\documentstyle{article}

\def\tfrac#1#2{{\textstyle {#1 \over #2}}}%
\def\dfrac#1#2{{\displaystyle {#1 \over #2}}}%
\def\stackunder#1#2{\mathrel{\mathop{#2}\limits_{#1}}}%

\begin{document}

\baselineskip = 18pt

\author{A.Y. Shiekh \\
{\it International Centre for Theoretical Physics, Miramare, Trieste, Italy}}
\title{{\bf Quantum Canonical Transformations revisited}}
\date{\vskip -4cm \rightline{\bf
IC/94/371, hep-th/9411199} \vskip 3cm}
\maketitle

\begin{abstract}
A preferred form for the path integral discretization is suggested that
allows the implementation of canonical transformations in quantum theory.
\end{abstract}

\section{Canonical Transformations in Quantum Theory}

Canonical transformations are of great utility in classical theory
[Goldstein, 1980]; however they do not work so well in quantum theory,
giving rise to anomalous (order $\hbar ^2$) potential like terms [Gervais
and Jevicki, 1980; Klauder, 1980]. Despite the age of the problem of quantum
canonical transformations, it continues to inspire interest [Anderson, 1993;
1993; 1994; 1994; Swanson, 1994].

Although equally applicable to quantum field theory, there is no virtue in
working within its added complexity, and the problem is discussed in the
context of quantum mechanics. This issue is well investigated in the setting
of the Hamiltonian (phase space) path integral formulation of quantum theory
[Feynman, 1948; 1965; Fanelli, 1976], due to its closeness to classical
theory with its commuting variables. A transition amplitude is then
generally written as:

\begin{equation}
\left\langle q_b,t_b\mid q_a,t_a\right\rangle =\int\limits_{-\infty }^\infty
Dq\int\limits_{-\infty }^\infty \dfrac{Dp}{2\pi \hbar }\exp \left( \dfrac
i\hbar \int\limits_{t_a}^{t_b}(p\dot q-H(q,p,t))dt\right)
\end{equation}

which is formal because it actually depends upon how it is discretized. In
general, each integral of the discretized path integral has a leading error
of order $\Delta t$, and this error is not lost in the limit of infinite
time refinement because there are $\frac
T{\Delta t}$ integrals ($T\equiv t_b-t_a$). So
this error makes a finite contribution in the
final limit, and it is exactly in this
sensitivity to $\Delta t$ that the
discretization scheme and equivalently the
operator ordering of the operator formalism
expresses itself [Schulman, 1981; Mayes and
Dowker, 1972; 1973]. However, if one could
locate a scheme with error of higher order than
$\Delta t$, no contribution would be hidden,
and one might anticipate a better behaved
object under formal manipulations. The idea is
that in moving into this scheme from another,
the hidden error would be exposed and correctly
accommodated within the proposed canonical
transformation.

For an investigation of the error, end point integrations can be neglected,
since the abandonment of any {\it finite} number of $\Delta t$ corrections
will make no
difference to the total error in the small time limit. Advantage will be
taken of this simplification throughout.

Although there will be a formal investigation, it might be
productive to first embark upon some numerical experiments to show the
existence of the types of objects sought.

\section{A `numerical' Investigation}

In the normal scheme, the standard path integral discretization is
given by a variety of non-equivalent forms, a typical example with end
points $(q_a,p_a)$, $(q_b,p_b)$ being:

\begin{equation}
I_\infty =\stackunder{N\to \infty }{\lim }\int\limits_{-\infty }^\infty
\ldots \int\limits_{-\infty }^\infty \ldots
\prod\limits_{j=1}^{N-1}dq_j\prod\limits_{i=1}^N\frac{dp_i}{2\pi \hbar
}{
\exp }\left( \frac i\hbar \sum_{k=1}^N\left(
p_k(q_k-q_{k-1})-H(q_{k-1},p_k)\Delta t\right) \right)
\label{eqn0}
\end{equation}

\rightline{ {\it where} $\Delta t=\frac TN$ }

It is well known, and will be confirmed below (both numerically and
analytically), that each of the above integrals has an error of order
$
\Delta t$. But just as the definition of the normal derivative has various
forms of differing error (of relevance to computer numerics), namely:

\begin{equation}
\frac{dx}{dt}\equiv \lim_{\Delta t\to 0}\frac{x(t+\Delta t)-x(t)}{\Delta t}
\end{equation}
or
\begin{equation}
\frac{dx}{dt}\equiv \lim_{\Delta t\to 0}\frac{x(t+\Delta t)-x(t-\Delta
t)}{
\Delta t}
\end{equation}

the first being accurate to order $\Delta
t$, and the second to order
$ (\Delta t)^2$; so one might also seek a
path integral scheme with higher order
error. There is a candidate in the more
symmetric (mid-point) form given by
[Shiekh, 1988; Klauder, 1980; Daubechies
and Klauder, 1985]:

\begin{equation}
\bar I_\infty =\stackunder{M\to \infty }{\lim }\int\limits_{-\infty }^\infty
\ldots \int\limits_{-\infty }^\infty \ldots
\prod\limits_{j=1}^{2M}dq_j\prod\limits_{i=0}^{2M+1}\frac{dp_i}{4\pi \hbar
}{
\exp }\left( \frac i\hbar \sum_{k=1}^{2M}\left(
p_k\frac{q_{k+1}-q_{k-1}} 2-H(q_k,p_k,t_k)\Delta t\right) \right)
\end{equation}

\rightline{ {\it where} $\Delta t=\frac T{2M+1}$ }

This symmetric discretization for the path integral exists only for even
subdivisions, and the first three time subdivision refinements are listed in
the appendix, where Wick rotation ($t\to -it$) has been performed to improve
numerical convergence, having set $\hbar =1$ and dropped the end
integrations, an act which does not effect the error contribution in the
small time limit, but simplifies matters.

The traditional form (equation \ref{eqn0}) is known to have an error of
order $\Delta t$, while
the symmetric form might be anticipated to have a higher order error. This
might first be investigated numerically before proceeding analytically, with
all end points held to zero (a coherent state like path integral).

These integrals differ from the limit by $\smallint =\smallint _\infty
+\alpha (\Delta t)^n+\ldots $, so (adopting the notation where the subscript
on $I$ indicates the number of $q$ (or $p$) integrations):

\begin{equation}
\smallint _2=\smallint _\infty +\alpha ({\frac 13}T)^n+\ldots
\end{equation}

\begin{equation}
\smallint _4=\smallint _\infty +\alpha ({\frac 15}T)^n+\ldots
\end{equation}

\begin{equation}
\smallint _6=\smallint _\infty +\alpha ({\frac 17}T)^n+\ldots
\end{equation}

Eliminating $\smallint _\infty $ and $\alpha $ in order to isolate the
leading order error yields:

\begin{equation}
\frac{\smallint _6-\smallint _4}{\smallint _4-\smallint _2}=\frac{
7^{-n}-5^{-n}}{5^{-n}-3^{-n}}
\end{equation}

For a leading order error of $(\Delta t)^1$ one would get a ratio of ${\frac
37}$ for small evolution times, while for $(\Delta t)^2$ one would get a
value of ${{\frac{{27}}{{98}}}}$. Using the trial Hamiltonian of a simple
harmonic oscillator, namely:

\begin{equation}
H={\tfrac 12}(p^2+q^2)
\end{equation}

avoids the heavy numerical work involved in the accurate evaluation of a
twelve dimensional integration (in fact, the integrals were performed
analytically with a computer mathematics package). Using a short evolution
time of $T=0.1$, and fixing the end points at zero, leads to the result for
the normal path integral:

\begin{equation}
\frac{I_6-I_4}{I_4-I_2}={\frac{{.9978625573\ldots -.9980047869\ldots
}}{{ .9980047869\ldots -.9983368743\ldots }}}=.42829\ldots
\end{equation}

the ${\frac 37}$ ($.42857...$) confirming that the leading order error is
indeed $(\Delta t)^1$; while for the symmetric form one gets:

\begin{equation}
\frac{\overline{I}_6-\overline{I}_4}{\overline{I}_4-\overline{I}_2}={\frac{{
.9951226146\ldots -.9952203945\ldots }}{{.9952203945\ldots
-.9955752212\ldots }}}=.27557\ldots
\end{equation}

the ${{\frac{{27}}{{98}}}}$ ($.27551...$) corresponding to a leading order
error of $(\Delta t)^2$, so confirming the suspicion that a higher order
scheme will be free of hidden contributions.

\section{Stochastic Terms}

The chance is taken here to derive the well-known results that in the path
integral $p\sim (\Delta t)^{-\frac 12}$ and $\Delta q\sim (\Delta t)^{\frac
12}$. An important exception to this rule is derived below.

Beginning from the Hamiltonian path integral, which consists of many
integrals of the form:

\begin{equation}
\int\limits_{-\infty }^\infty dq\int\limits_{-\infty }^\infty dp\exp \left(
\frac i\hbar \left( p\Delta q-H(q,p,t)\Delta t\right) \right)
\end{equation}

It is assumed here that one is not dealing with unphysical, higher
derivative theories
(which would demand higher order `momenta') so that the momentum is not
found higher than quadratic order or negative powers. As a result the
canonical transformation
is somewhat limited in that it should not map to a theory of higher order,
and the integral becomes:

\begin{equation}
\int\limits_{-\infty }^\infty dq\int\limits_{-\infty }^\infty dp\exp
 \left(
\frac i\hbar \left( p\Delta q
-\left( \frac{p^2}{2m(q,t)} + \gamma(q,t) p + V(q,t) \right) \Delta t
\right) \right)
\label{eqn1}
\end{equation}

The $p$ integral may then be performed using the Gaussian result:

\begin{equation}
\int\limits_{-\infty }^\infty
e^{- \alpha s^2 - \beta s} ds
=
\sqrt{\frac{\pi}{\alpha} } e^{\frac{\beta^2}{4\alpha}}
\end{equation}

Now since the $p^2$ generates a
$1 / \alpha$, where $\alpha =
i \Delta t / {2m} \hbar$, so each $p$ (for
even powers) contributes like $ (\Delta
t)^{-\frac 12}$, while $p$ alone generates
a $\beta^2$, i.e. $-(\gamma \Delta t-\Delta q)^2/ \hbar^2$,
so that $p$ in odd powers contributes like
$(\Delta t)^1$. In
performing the $p$ integrals in equation
\ref{eqn1} and so obtaining the Lagrange
formalism; $p$ becomes
$m(\frac{\Delta q}{\Delta t} - \gamma)$, so that each
$ \Delta q\sim (\Delta t)^{\frac 12}$ [c.f. $p\sim (\Delta t)^{-\frac 12}$] in
even powers, and $\sim (\Delta t)^1$ for odd powers.
Another way to see this result is to expand out the $p$ term, namely
$\exp(ip(\Delta q-\gamma \Delta t)/ \hbar)$, and then note from
symmetry that the contribution starts only at the second term, i.e.
for $p^2 (\Delta q-\gamma \Delta t)^2$, which indeed contributes like
$(\Delta t)^1$. This higher order contribution for odd powers will be
crucial later.

It is in this way that the contributing class of paths are seen to be
stochastic (or Brownian) in nature. This behaviour of the path
integral must be carefully accounted for when working to order $\Delta t$.

\section{Canonical Transformations in the Symmetric Path Integral}

In classical mechanics a canonical transformation is one that preserves the
least action principle [Goldstein, 1980]. For the path integral one might
analogously require that there be a path integral representation in the new
variables $(Q,P,t)$, if one existed in the old ones $(q,p,t)$. The fact that
the symmetric path integral has no `hidden' parts should guarantee that
formal canonical transformations are now valid. This in explicitly
demonstrated below.

A canonical transformation should be system independent, that is to say, the
transformation should be canonical not only for some specific system, but
for all problems with the same degrees of freedom. The amplitude may alter
under such a transformation by at most a phase factor. So {\it formally}
one gets:

\begin{equation}
\left.
\begin{array}{c}
\int\limits_{-\infty }^\infty Dq\int\limits_{-\infty }^\infty
\dfrac{Dp}{ 2\pi \hbar }\exp \left( \dfrac i\hbar
\int\limits_{t_a}^{t_b}\left( p\dot q-H(q,p,t)\right) dt\right) \\
=\exp \left( \frac i\hbar (F_b-F_a)\right) \int\limits_{-\infty }^\infty
DQ\int\limits_{-\infty }^\infty \dfrac{DP}{2\pi \hbar }\exp \left( \dfrac
i\hbar \int\limits_{t_a}^{t_b}\left( P\dot Q-K(Q,P,t)\right) dt\right)
\end{array}
\right.  \label{eqn2}
\end{equation}

with $F(q,Q,t)$ being an arbitrary smooth function. Since the above
equation is to
be true for all Hamiltonians, one gets:

\begin{equation}
p\dot q-H=P\dot Q-K+\frac{dF}{dt}
\end{equation}

the same condition as in classical mechanics, with $F$ the generating
function of the canonical transformation. For $F=F(q,Q,t)$ one gets:

\begin{equation}
p\dot q-H=P\dot Q-K+\left. \frac{\partial F}{\partial q}\right| _{Q,t}\dot
q+\left. \frac{\partial F}{\partial Q}\right| _{q,t}\dot Q+\left.
\frac{
\partial F}{\partial t}\right| _{q,Q}
\end{equation}

from which follows, by the independence of $q$ and $Q$:

\begin{equation}
p=\left. \frac{\partial F}{\partial q}\right| _{Q,t}
\end{equation}

\begin{equation}
P=\left. {-}\frac{\partial F}{\partial Q}\right| _{q,t}
\end{equation}

\begin{equation}
K=H+\left. \frac{\partial F}{\partial t}\right| _{q,Q}
\end{equation}

All this work was formal, and now one is ready to apply this machinery to
the symmetric discretization of the path integral in the hope that there
will be no corrections. The formal canonical transformation of equation \ref
{eqn2} now becomes (ignoring end integrals):

\begin{equation}
\left.
\begin{array}{c}
\stackunder{M\to \infty }{\lim }\int\limits_{-\infty }^\infty
..\int\limits_{-\infty }^\infty
..\prod\limits_{j=1}^{2M}dq_j\prod\limits_{i=1}^{2M}\dfrac{dp_i}{4\pi \hbar
} {\exp }\left( \dfrac i\hbar
\sum\limits_{k=1}^{2M}p_k\dfrac{q_{k+1}-q_{k-1}}
2-H(q_k,p_k,t_k)\Delta t\right) \\  {}{}{\Rightarrow }\stackunder{M\to
\infty }{\lim }\int\limits_{-\infty }^\infty ..\int\limits_{-\infty
}^\infty
..\prod\limits_{j=1}^{2M}dQ_j\prod\limits_{i=1}^{2M}\dfrac{dP_i}{4\pi
\hbar } {\exp }\left( \dfrac i\hbar
\sum\limits_{k=1}^{2M}P_k\dfrac{Q_{k+1}-Q_{k-1}}
2-K(Q_k,P_k,t_k)\Delta t+\Delta F\right)
\end{array}
\right.
\end{equation}

\rightline{
{\it where:} $\Delta F\equiv \dfrac{
F(q_{k+1},Q_{k+1},t_{k+1})-F(q_{k-1},Q_{k-1},t_{k-1})}2$
}

Now, $F$ itself, like all other terms in the action, cannot be worse that
$ (\Delta t)^0$ in strength, for the same reason given before that
`higher derivative' actions are excluded from this discussion. By
expanding $F$ and also using the facts that $\Delta q\sim (\Delta
t)^{\frac 12}$ and $\Delta Q\sim (\Delta t)^{\frac 12}$ for even
powers, and
$\sim (\Delta t)^1$
 for odd powers, one gets
the crucial result:

\begin{equation}
\Delta F=\left. \frac{\partial F}{\partial Q}\right| _{q,t}\Delta Q+\left.
\frac{\partial F}{\partial q}\right| _{Q,t}\Delta q+\left. \frac{\partial
F}{
\partial t}\right| _{q,Q}\Delta t+O(\Delta t)^{\frac 32}
\end{equation}

Had one not used the symmetric scheme, anomalous (order $\hbar ^2$) terms
would have entered here with the even powers of $\Delta q$ and $\Delta Q$
then present\footnote{It was the lack of a viscous
term in our numerical experiment that
lead to a $(\Delta t)^2$ error, as
opposed to the more general
$(\Delta t)^{\frac 32}$ error.}. Dropping the $(\Delta t)^{\frac 32}$ term in
the above  (it is a $\Delta t$ term that does not disappear in the limit)
leads to:

\begin{equation}
\Delta F=-P\Delta Q+p\Delta q+\left. \frac{\partial F}{\partial t}\right|
_{q,Q}\Delta t
\end{equation}

and from this one gets the anomaly free transformation of $p\Delta q$,
namely:

\begin{equation}
p\Delta q=P\Delta Q+\Delta F-\left. \frac{\partial F}{\partial t}\right|
_{q,Q}\Delta t
\end{equation}

The Jacobian for the measure is unity, and the Hamiltonian conversion is
equally trivial, since they are both local, and leads to:

\begin{equation}
H=K+\left. \frac{\partial F}{\partial t}\right| _{q,Q}
\end{equation}

Putting this all together leads to the sought after equality, namely:

\begin{equation}
\left.
\begin{array}{c}
\stackunder{M\to \infty }{\lim }\int\limits_{-\infty }^\infty \ldots
\int\limits_{-\infty }^\infty \ldots
\prod\limits_{j=1}^{2M}dq_j\prod\limits_{i=0}^{2M+1}\dfrac{dp_i}{4\pi \hbar
}{
\exp }\left( \dfrac i\hbar
\sum\limits_{k=1}^{2M}p_k\dfrac{q_{k+1}-q_{k-1}}
2-H(q_k,p_k,t_k)\Delta t\right) = \\  {\exp }\left( \dfrac i\hbar
(F_b-F_a)\right) \stackunder{M\to \infty }{\lim }
\int\limits_{-\infty }^\infty \ldots \int\limits_{-\infty }^\infty \ldots
\prod\limits_{j=1}^{2M}dQ_j\prod\limits_{i=0}^{2M+1}\dfrac{dP_i}{4\pi \hbar
}{
\exp }\left( \dfrac i\hbar
\sum\limits_{k=1}^{2M}P_k\dfrac{Q_{k+1}-Q_{k-1}}
2-K(Q_k,P_k,t_k)\Delta t\right)
\end{array}
\right.
\end{equation}

having restored the end integrals (at the cost of no error in the limit).

This completes the demonstration that the symmetric form canonically
transforms cleanly.

\section{Corollaries}

The utility of having a canonically invariant prescription for the path
integral extends beyond just canonical transformations. It permits a
consistent (although not unique) quantization of a classical system
[Chernoff, 1981; Kapoor, 1984; Dirac, 1925; 1958]. Such consistency is part
of the way to making the path integral a well-defined object [Daubechies and
Klauder, 1985]. The phase space path integral also has a sensitivity to the
order in which the $p$ and $q$ integrations are performed. This problem and
its solution is discussed elsewhere
[Shiekh, 1990].

It should be emphasised that the symmetric path integral is favoured (not
compelled) over other prescriptions, in that it generates no anomalous
additional terms during the canonical transformation. Note, however, that to
get into and from this scheme, stochastic ($\hbar ^2$) terms appear. The
virtue of no terms occurring during the transformation is that the
Hamiltonian behaves well under the transformation, and is then (for example)
trivialised by a Hamilton-Jacobi transformation.

\section{Acknowledgments}

I should like to thank ICTP for support during this work.

\section{Appendix: Explicit Discretizations}

The first three (Wick rotated) symmetric path integral discretizations with
$
\hbar =1$ and end points held fixed (coherent state like path integral) are
given by:

\begin{equation}
\bar I_2={\frac 1{{(4\pi )^2}}}\int\limits_{-\infty }^\infty
dq_1dq_1\int\limits_{-\infty }^\infty dp_2dp_2\exp \left(
\begin{array}{c}
\frac i2(p_1(q_2-q_a)){-}H(q_1,p_1,t_1)T/3 \\
{+}\frac i2(p_2(q_b-q_1)){-}H(q_2,p_2,t_2)T/3
\end{array}
\right)
\end{equation}

\begin{equation}
\bar I_4={\frac 1{{(4\pi )^4}}}\int\limits_{-\infty }^\infty
dq_1..dq_4\int\limits_{-\infty }^\infty dp_1..dp_4\exp \left(
\begin{array}{c}
\frac i2(p_1(q_2-q_a)){-}H(q_1,p_1,t_1)T/5 \\
{+}\frac i2(p_2(q_3-q_1)){-}H(q_2,p_2,t_2)T/5 \\
{+}\frac i2(p_3(q_4-q_2)){-}H(q_3,p_3,t_3)T/5 \\
{+}\frac i2(p_4(q_b-q_3)){-}H(q_4,p_4,t_4)T/5
\end{array}
\right)
\end{equation}

\begin{equation}
\bar I_6={\frac 1{{(4\pi )^6}}}\int\limits_{-\infty }^\infty
dq_1..dq_6\int\limits_{-\infty }^\infty {\,}dp_1..dp_6\exp \left(
\begin{array}{c}
\frac i2(p_1(q_2-q_a)){-}H(q_1,p_1,t_1)T/7 \\
{+}\frac i2(p_2(q_3-q_1)){-}H(q_2,p_2,t_2)T/7 \\
{+}\frac i2(p_3(q_4-q_2)){-}H(q_3,p_3,t_3)T/7 \\
{+}\frac i2(p_4(q_5-q_3)){-}H(q_4,p_4,t_4)T/7 \\
{+}\frac i2(p_5(q_6-q_4)){-}H(q_5,p_5,t_5)T/7 \\
{+}\frac i2(p_6(q_b-q_5)){-}H(q_6,p_6,t_6)T/7
\end{array}
\right)
\end{equation}

These might be compared against the corresponding, more usual,
discretizations given by:

\begin{equation}
I_2={\frac 1{{(2\pi )^2}}}\int\limits_{-\infty }^\infty
dq_1dq_1\int\limits_{-\infty }^\infty dp_2dp_2\exp \left(
\begin{array}{c}
ip_1(q_1-q_a)-H(q_a,p_1)T/3 \\
{+}ip_2(q_2-q_1)-H(q_1,p_2)T/3 \\
{+}ip_b(q_b-q_2)-H(q_2,p_b)T/3
\end{array}
\right)
\end{equation}

\begin{equation}
I_4={\frac 1{{(2\pi )^4}}}\int\limits_{-\infty }^\infty
dq_1..dq_4\int\limits_{-\infty }^\infty dp_1..dp_4\exp \left(
\begin{array}{c}
ip_1(q_1-q_a)-H(q_a,p_1)T/5 \\
{+}ip_2(q_2-q_1)-H(q_1,p_2)T/5 \\
{+}ip_3(q_3-q_2)-H(q_2,p_3)T/5 \\
{+}ip_4(q_4-q_3)-H(q_3,p_4)T/5 \\
{+}ip_b(q_b-q_4)-H(q_4,p_b)T/5
\end{array}
\right)
\end{equation}

\begin{equation}
I_6={\frac 1{{(2\pi )^6}}}\int\limits_{-\infty }^\infty
dq_1..dq_6\int\limits_{-\infty }^\infty {\,}dp_1..dp_6\exp \left(
\begin{array}{c}
ip_1(q_1-q_a)-H(q_a,p_1)T/7 \\
{+}ip_2(q_2-q_1)-H(q_1,p_2)T/7 \\
{+}ip_3(q_3-q_2)-H(q_2,p_3)T/7 \\
{+}ip_4(q_4-q_3)-H(q_3,p_4)T/7 \\
{+}ip_5(q_5-q_4)-H(q_4,p_5)T/7 \\
{+}ip_6(q_6-q_5)-H(q_5,p_6)T/7 \\
{+}ip_b(q_b-q_6)-H(q_6,p_b)T/7
\end{array}
\right)
\end{equation}

\section{References}

H. Goldstein, {\it `Classical Mechanics'}, (Addison-Wesley, Reading, MA,
2nd ed, (1980).

J. Gervais and A. Jevicki, {\it ``Point canonical transformations in the
path integral''}, Nucl. Phys. B {\bf 10}, (1980), 93-.

J. Klauder, {\it `Path Integrals'}, Proceedings of the XIX. Internationale
Universit\"atswochen f\"ur Kernphysik, Schladming, Austria, Acta Physica
Austriaca, {\bf Suppl. XXII}, (1980), 3-.

A. Anderson, {\it `Quantum Canonical Transformations: Physical Equivalence
of Quantum Theories'}, Phys. Lett. B {\bf 305}, (1993), 67-.

A.Anderson, {\it `Quantum Canonical Transformations and Integrability:
Beyond Unitarity Transformations'}, Phys. Lett. B {\bf 319}, (1993),
157-.

A.Anderson, {\it `Canonical Transformations in Quantum Mechanics'},
Ann. Phys. (NY) {\bf 232}, (1994), 292-.

A. Anderson, {\it `Special Functions from Canonical Transformations'},
J. Math. Phys. {\bf 35}, (1994), 6018-.

M. Swanson, {\it `Canonical Transformations and Path Integral Measures'},
to appear in Phy. Rev. A, (1994).

R. Feynman, {\it `Space-Time Approach to Quantum Mechanics'}, Rev. Mod.
Phys. {\bf 20} (1948), 267-. (reprinted in {\it ``Quantum
Electrodynamics''}, Ed. J.Schwinger, Dover, 1958.)

R. Feynman and A. Hibbs, {\it ``Quantum Mechanics and Path Integrals''},
McGraw-Hill, (1965).

R. Fanelli, {\it `Canonical Transformations and Phase Space Path
Integrals'}, J. Math. Phys., {\bf 17}, (1976), 490-

L. Schulman, {\it ``Techniques and Applications of Path Integration''},
Wiley, (1981).

I. Mayes and J. Dowker, {\it `Canonical Functional Integrals in General
Coordinates'}, Proc. R. Soc. Lond., A.{\bf 327}, (1972), 131-.

I. Mayes and J. Dowker, {\it `Hamiltonian Orderings and Functional
Integrals' }, J. Math. Phys., {\bf 14}, (1973), 434-.

A. Shiekh, {\it ``Canonical transformations in quantum mechanics: A
canonically invariant path integral''}, J. Math. Phys., {\bf 29} (4),
(1988), 913-.

I. Daubechies and J. Klauder, {\it `True Measures for real time Path
Integrals'}, in {\it ``Path Integrals from meV to MeV''}, Ed. M. Gutzwiller,
A. Inomata, J. Klauder and L. Streit, World Scientific, (1985), 425-.

A. Shiekh, {\it ``The trivialization of constrains in quantum theory
(working in a general gauge/parametrization)''}, J. Math. Phys., {\bf 31}
(1), (1990), 76-.

P. Chernoff, {\it `Mathematical Obstructions to Quantization'}, Had. J.,
{\bf 4}, (1981), 879-.

A. Kapoor, {\it `Quantization in Nonlinear Coordinates via Hamiltonian Path
Integrals'}, Phys.Rev., D {\bf 29}, (1984), 2339-.

P. Dirac, {\it `The Fundamental Equations of Quantum Mechanics'}, Proc. Roy.
Soc. A {\bf 109}, (1925), 642-.

P. Dirac, {\it ``The Principles of Quantum Mechanics''}, 4th Ed., Oxford,
(1958).

\end{document}